\documentstyle[11pt,epsfig]{article}

\def\appendix{\par
 \setcounter{section}{0}
 \setcounter{subsection}{0}
 \def\thesection{Appendix \Alph{section}}
 \def\theequation{\Alph{section}.\arabic{equation}}
 \setcounter{equation}{0}}

\def\le{\left(}
\def\ri{\right)}

\def\no{\nonumber}

\def\f12{\frac{1}{2}}

\def\pd{\partial}

\begin{document}
\begin{titlepage}
\vskip 2cm
\begin{center}
{\Large \bf Triangle UD integrals in the position space}\\
\vskip 1cm  
Igor Kondrashuk$^{a}$ and Anatoly Kotikov$^{b,c}$\\
\vskip 5mm  
{\it  (a) Departamento de Ciencias B\'asicas, \\
Universidad del B\'\i o-B\'\i o, Campus Fernando May, Casilla 447, Chill\'an, Chile} \\
\vskip 1mm
{\it  (b) Bogoliubov Laboratory of Theoretical Physics, Joint Institute for Nuclear Research, \\
Dubna, Russia} \\
\vskip 1mm
{\it  (c) II. Institut fur Theoretische Physik, Universitat Hamburg, \\
Luruper Chaussee 149, 22761 Hamburg, Germany} \\
\end{center}
\vskip 20mm
\begin{abstract}
We investigate triangle UD ladder integrals in the position space. The investigation is necessary to find an all-order in loop solution 
for an auxiliary $Lcc$ correlator in Wess-Zumino-Landau gauge of the maximally supersymmetric Yang-Mills theory
and to present correlators of dressed mean gluons in terms of it in all loops. We show that triangle UD ladder diagrams 
in the position space can be expressed in terms of the same UD functions $\Phi^{(L)},$  in terms of which they were represented in the 
momentum space, for an arbitrary number of rungs. 
\end{abstract}
\end{titlepage}

As has been shown in Refs. \cite{Cvetic:2004kx} - \cite{Kondrashuk:2000qb}, Slavnov-Taylor (ST) identity predicts that the correlators 
of dressed mean fields for ${\cal N}= 4$ supersymmetric Yang-Mills theory in {\em the position space} can be represented 
in terms of Usyukina-Davydychev (UD) integrals (at least at the two loop planar level). Indeed, at that level the auxiliary $Lcc$ correlator in the position space 
in Wess-Zumino-Landau gauge of the maximally supersymmetric Yang-Mills theory is a function of Davydychev integral $J(1,1,1)$ 
\cite{Cvetic:2006iu,Cvetic:2007fp,Cvetic:2007ds} which is the first integral in the chain of UD integrals 
\cite{Davydychev:1992xr,Usyukina:1992jd,Usyukina:1993ch}. By using ST identity, we can express all the correlators in terms of this correlator 
in that theory. Furthermore, using the method of Ref. \cite{Cvetic:2007ds}, one can expect that at the higher loop orders 
in the position space the triangle UD ladder integral contributions to that auxiliary correlator will survive only.  
Strictly speaking, the powers of d`Alambertian applied to the $L$-field vertex of the triangle ladder  will contribute only. 
In this paper we show that such constructions are the UD functions of the spacetime intervals. Conformal invariance of the effective 
action of dressed mean fields in the position space, suggested in  Refs. \cite{Cvetic:2004kx,Kondrashuk:2004pu,Kang:2004cs}, 
corresponds to the property of conformal invariance of the UD functions in the position space.

The UD integrals correspond to the momentum representation of three-point ladder diagrams (triangle ladders) and four-point ladder 
diagrams and were defined and calculated in Refs. \cite{Usyukina:1992jd,Usyukina:1993ch} in {\em the momentum space}, 
and the result can be written in terms of the UD functions $\Phi^{(L)}$ of conformally invariant ratios of momenta 
\footnote{In the position space Feynman diagrams contain integrations over coordinates of internal vertices. 
Integration over internal vertices appears in dual representation of the momentum diagrams too \cite{Kazakov:1986mu,Kazakov:1987jk}}. 
In the momentum space it was shown that  the UD functions are the only contributions (at least up to three loops) 
to off-shell four-point correlator of gluons that corresponds to four gluon amplitude \cite{Drummond:2006rz,Bern:2005iz}. 
The conformal invariance of UD functions was used in the momentum space to calculate four-point amplitude and 
to classify all possible contributions to it \cite{Bern:2006ew,Nguyen:2007ya}. Later, the conformal symmetry in the momentum space  
appeared on the string side in the Alday-Maldacena approach \cite{Alday:2007hr} in the limit of strong coupling.

\begin{figure}[htb]
\begin{minipage}[h]{0.7\linewidth}
\centering\epsfig{file=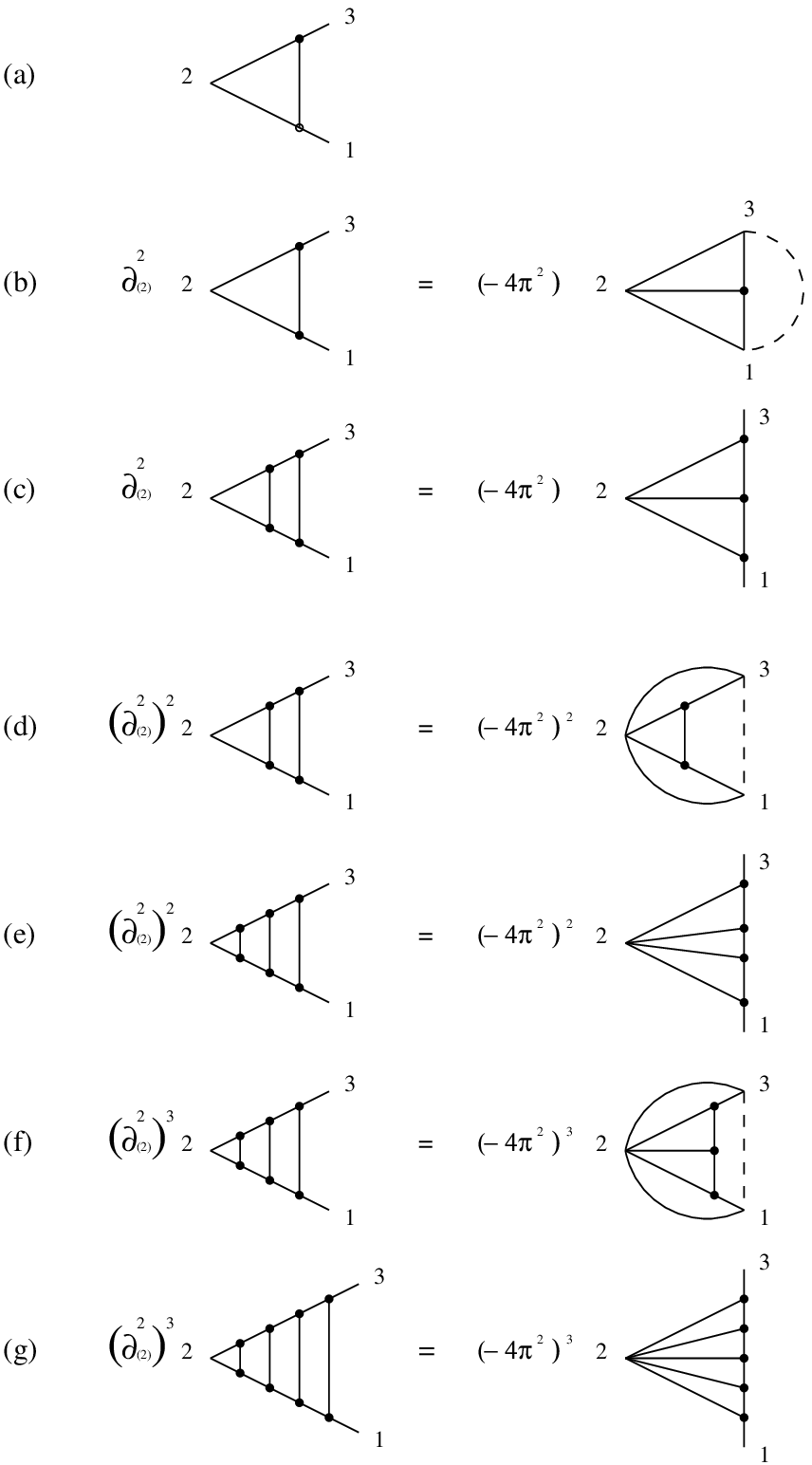,width=\linewidth}
\end{minipage}
\caption{\footnotesize Chain of transformations}
\label{Chain}
\end{figure}

In this paper we use two things known from the literature. These are the iterative definition of the UD functions, that is Eq. (23) of
Ref.\cite{Usyukina:1992jd}, and the dual graphical representation for four-point momentum UD integrals in the form of ``diamonds'' 
\cite{Isaev:2003tk,Drummond:2006rz}. Before starting the demonstration, we outline some basic points of it. In Ref. \cite{Kondrashuk:2008ec}
we have proved the identity \footnote{Our definition for UD functions is  $\Phi_{New}^{(L)} = (\pi^2)^L\Phi_{Old}^{(L)},$ where  $\Phi_{New}^{(L)}$ is  $\Phi^{(L)}$ of this paper,
and   $\Phi_{Old}^{(L)}$ is the original UD function $\Phi^{(L)}$ of Refs. \cite{Usyukina:1992jd,Usyukina:1993ch}.}
\begin{eqnarray}
\int~d^4y~d^4z \frac{1}{[2y][1y][3z][yz][2z]} = \frac{1}{[31]} \Phi^{(2)}\le \frac{[12]}{[31]},\frac{[23]}{[31]}\ri \label{f-i} 
\end{eqnarray}
by conformal transformation of the integrand. 
The l.h.s. of this relation corresponds to the l.h.s. of the line $(a)$ of Fig.(\ref{Chain}). In this paper we assume the notation of Ref. \cite{Cvetic:2006iu}, 
where $[Ny]= (x_N - y)^2$ and analogously for $[Nz]$ and $[yz],$ that is, $N=1,2,3$ stands for $x_N=x_1,x_2,x_3,$ respectively, throughout all the paper. In the momentum space, 
Eq.(\ref{f-i}) might be understood as a relation between the momentum integrals that correspond
 to the l.h.s. and the r.h.s. of  Fig.(2) of Ref. \cite{Usyukina:1994iw} which was derived by using the trick of integration by parts. However, in 
Ref. \cite{Usyukina:1994iw} the position space picture was not analysed and it was not shown that Fourier transform of the second UD integral is 
the same integral \cite{Kondrashuk:2008ec}. Here we demonstrate the validity of a similar property for any UD function. 

The line $(b)$ of Fig.(\ref{Chain}) is Fig.(5) of Ref.\cite{Kondrashuk:2008ec}. It corresponds to the equation  
\begin{eqnarray}
\pd_{(2)}^2 \int~Dy~Dz \frac{1}{[2y][1y][3z][yz][2z]} = -\frac{4[31]}{[12][23]}~J(1,1,1). \label{BB1}
\end{eqnarray}
This equation has been generated by the computer program of Ref. \cite{Cvetic:2007ds} by making use of formulas of 
Refs. \cite{Cvetic:2006iu,Cvetic:2007fp} derived by Gegenbauer polynomial technique. Without modifications of the 
integral measure made in Ref. \cite{Cvetic:2006iu} 
this equation is \footnote{All internal vertices of the diagrams in this paper correspond to the standard four-dimensional integral measure.}
\begin{eqnarray}
\pd_{(2)}^2 \int~d^4y~d^4z \frac{1}{[2y][1y][3z][yz][2z]} = -\frac{4\pi^2[31]}{[12][23]}~J(1,1,1). \label{BB2}
\end{eqnarray}
In Ref. \cite{Eden:1998hh} this equation was obtained by direct differentiation of the second UD integral. It can be demonstrated in several ways.
For example, the direct differentiation of the r.h.s. of Eq. (\ref{f-i}) using the Eq.(23) of Ref.\cite{Usyukina:1992jd} and Eq.(12) 
of Ref.\cite{Usyukina:1993ch}  produces 
\begin{eqnarray}
\pd_{(2)}^2\frac{1}{[31]} \Phi^{(2)}\le \frac{[12]}{[31]},\frac{[23]}{[31]}\ri = \no\\
~[31]\pd_{(2)}^2~\frac{1}{[31]^2} \Phi^{(2)} \le \frac{[12]}{[31]},\frac{[23]}{[31]}\ri =  
~[31]\pd_{(2)}^2 C^{(2)}([12],[23],[31]) = \no\\
~[31] \pd_{(2)}^2 \int~d^4y~\frac{C^{(1)}([(12)+y],[(23)-y],[31])}{[(12)+y]~[(23)-y]~[y]} = \no\\
~[31] \pd_{(2)}^2 \int~d^4y~\frac{C^{(1)}([(12)-y],[(32)-y],[31])}{[(12)-y]~[(32)-y]~[y]} = 
~[31] \pd_{(2)}^2 \int~d^4y~\frac{C^{(1)}([1y],[3y],[31])}{[1y]~[2y]~[3y]} = \no\\
~ -4\pi^2[31] \int~d^4y~\delta(2y)\frac{C^{(1)}([1y],[3y],[31])}{[1y]~[3y]} = 
 -4\pi^2[31] \frac{C^{(1)}([12],[23],[31])}{[12]~[23]} = \no\\
- \frac{4\pi^2}{[12][23]}~\Phi^{(1)}\le\frac{[12]}{[31]},\frac{[23]}{[31]}\ri. \label{BB3} 
\end{eqnarray}
On the other hand, by using conformal transformation the r.h.s. of Eq.(\ref{BB2}) and Eq.(\ref{BB3}) can be related. 
Three-point UD functions can be transformed to four-point UD functions due to Jacobian of conformal transformation, since 
under this transformation each three-point internal vertex transforms to four-point internal vertex 
with a new leg growing from the internal vertex to the point $0$ which is the inition of the reference system. The conformal substitution for each vector of the 
integrand (including the external vectors) is 
\begin{eqnarray}
y_\mu = \frac{y'_\mu}{{y'}^2}, ~~~~ z_\mu = \frac{z'_\mu}{{z'}^2}, \label{CT} 
\end{eqnarray}
and in the simplest case of the first UD function we have  
\begin{eqnarray*}
J(1,1,1) = \int~d^4y~ \frac{1}{[1y][2y][3y]} = [1'][2'][3']\int~d^4y'\frac{1}{[1'y'][2'y'][3'y'][y']} =\\
~[1'][2'][3'] \frac{1}{[3'1'][2']} \Phi^{(1)}\le \frac{[1'2'][3']}{[3'1'][2']},\frac{[1'][2'3']}{[3'1'][2']}\ri  = \\
\frac{[1'][3']}{[3'1']} \Phi^{(1)}\le \frac{[1'2'][3']}{[3'1'][2']},\frac{[1'][2'3']}{[3'1'][2']}\ri  =  
\frac{1}{[31]} \Phi^{(1)}\le \frac{[12]}{[31]},\frac{[23]}{[31]}\ri .
\end{eqnarray*}

\begin{figure}[htb]
\begin{minipage}[h]{0.9\linewidth}
\centering\epsfig{file=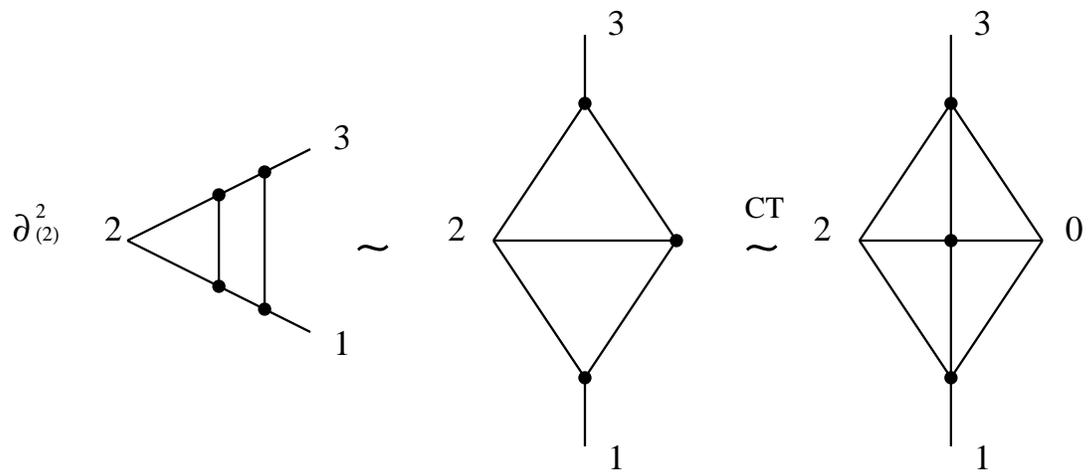,width=\linewidth}
\end{minipage}
\caption{\footnotesize Transformation of two rungs diagram}
\label{Two rungs}
\end{figure}

The line $(c)$ is the direct use of the line $(b)$ and, as  it can be proved  by the sequence of transformations depicted 
in Fig.(\ref{Two rungs}), its r.h.s. is proportional to $\Phi^{(3)},$ indeed 
\begin{eqnarray}
\int~d^4y~d^4z~d^4u \frac{1}{[2y][2u][2z][3z][1y][uz][uy]} = \no\\
~[2']^3[3'][1']\int~d^4y'~d^4z'~d^4u'\frac{1}{[2'y'][2'u'][2'z'][3'z'][1'y'][u'z'][u'y'][y'][z'][u']} = \no\\
~[2']^3[3'][1'] \frac{1}{[3'1'][2']^3} \Phi^{(3)}\le \frac{[1'2'][3']}{[3'1'][2']},\frac{[1'][2'3']}{[3'1'][2']}\ri  = \no\\
\frac{[3'][1']}{[3'1']} \Phi^{(3)}\le \frac{[1'2'][3']}{[3'1'][2']},\frac{[1'][2'3']}{[3'1'][2']}\ri  =  
\frac{1}{[31]} \Phi^{(3)}\le \frac{[12]}{[31]},\frac{[23]}{[31]}\ri . \label{BB4}
\end{eqnarray}
The factor $-4\pi^2$ on the r.h.s. of the line $(c)$ came from Eq. (\ref{BB3}).

The line $(d)$ can be obtained by the direct differentiation of the previous result,
\begin{eqnarray}
\pd_{(2)}^2\frac{1}{[31]} \Phi^{(3)}\le \frac{[12]}{[31]},\frac{[23]}{[31]}\ri = \no\\
~[31]^2\pd_{(2)}^2~\frac{1}{[31]^3} \Phi^{(3)} \le \frac{[12]}{[31]},\frac{[23]}{[31]}\ri =  
~[31]^2\pd_{(2)}^2 C^{(3)}([12],[23],[31]) = \no\\
~[31]^2 \pd_{(2)}^2 \int~d^4y~\frac{C^{(2)}([(12)+y],[(23)-y],[31])}{[(12)+y]~[(23)-y]~[y]} = \no\\
~[31]^2 \pd_{(2)}^2 \int~d^4y~\frac{C^{(2)}([(12)-y],[(32)-y],[31])}{[(12)-y]~[(32)-y]~[y]} = \no\\
~[31]^2 \pd_{(2)}^2 \int~d^4y~\frac{C^{(2)}([1y],[3y],[31])}{[1y]~[2y]~[3y]} = \no\\
~ -4\pi^2[31]^2 \int~d^4y~\delta(2y)\frac{C^{(2)}([1y],[3y],[31])}{[1y]~[3y]} = 
 -4\pi^2[31]^2 \frac{C^{(2)}([12],[23],[31])}{[12]~[23]} = \no\\
- \frac{4\pi^2}{[12][23]}~\Phi^{(2)}\le\frac{[12]}{[31]},\frac{[23]}{[31]}\ri. \label{BB5} 
\end{eqnarray}
Using this and taking into account the line $(a)$ and the result for it represented by Eq.(\ref{f-i}) we have obtained the r.h.s. of line $(d).$

The line $(e)$ is the direct use of  the line $(d).$ The r.h.s. of the line  $(e)$  is proportional to 
\begin{eqnarray}
\frac{1}{[31]} \Phi^{(4)}\le \frac{[12]}{[31]},\frac{[23]}{[31]}\ri. \label{BB6}
\end{eqnarray}
The proof  of this statement repeats proof (\ref{BB4}), the only difference is that instead of three internal vertices between the points 1 and 3 on the r.h.s. of the line $(c),$ 
we have  four internal vertices for the r.h.s. of the line $(e).$

The line $(f)$ is the repetition of the trick of Eq. (\ref{BB3}) and Eq. (\ref{BB5}) with the lines $(b)$ and $(d).$ Indeed, applying d'Alambertian to Eq.(\ref{BB6}), we obtain 
\begin{eqnarray}
\pd_{(2)}^2\frac{1}{[31]} \Phi^{(4)}\le \frac{[12]}{[31]},\frac{[23]}{[31]}\ri = - \frac{4\pi^2}{[12][23]}~\Phi^{(3)}\le\frac{[12]}{[31]},\frac{[23]}{[31]}\ri. \label{BB7} 
\end{eqnarray}
Using this and taking into account the line $(c)$ and the result for its r.h.s. represented by Eq.(\ref{BB4}) we have obtained the r.h.s. of the line $(f).$

The line $(g)$ is the direct use of line $(f).$  Repeating the proof of Eq.(\ref{BB4}) we obtain that the r.h.s. of the line $(g)$ is proportional to 
\begin{eqnarray*}
\frac{1}{[31]} \Phi^{(5)}\le \frac{[12]}{[31]},\frac{[23]}{[31]}\ri. 
\end{eqnarray*}

We can proceed this chain of constructions to an arbitrary number of rungs, and analysing the previous results and Fig.(\ref{Chain}), we obtain for $n$-rungs triangle 
UD ladder diagram $T_n([12],[23],[31])$ in the position space the following relations
\begin{eqnarray*}
\le \pd_{(2)}^2 \ri^{n-1}~T_n([12],[23],[31]) = \frac{(-4\pi^2)^{n-1}}{[31]} \Phi^{(n+1)}\le \frac{[12]}{[31]},\frac{[23]}{[31]}\ri, \\
\le \pd_{(2)}^2 \ri^{n}~T_n([12],[23],[31]) = \frac{(-4\pi^2)^n}{[12][23]} \Phi^{(n)}\le \frac{[12]}{[31]},\frac{[23]}{[31]}\ri, \\
\le \pd_{(2)}^2 \ri^{n}~T_{n+1}([12],[23],[31]) = \frac{(-4\pi^2)^n}{[31]} \Phi^{(n+2)}\le \frac{[12]}{[31]},\frac{[23]}{[31]}\ri.
\end{eqnarray*}

These relations show that the auxiliary $Lcc$ correlator in the position space in the maximally supersymmetric Yang-Mills theory 
can be represented in all loops in terms of the UD functions. Indeed, according to the technique developed in Ref. \cite{Cvetic:2007ds},
any contribution to the correlator can be expressed in terms of powers of d'Alambertian applied to a leg of scalar integrals.
Using the graphical identity of Refs.\cite{Kondrashuk:2008ec,Usyukina:1994iw} any four-point internal vertex of those scalar integrals can be 
presented in terms of three-point internal vertices. Other distributions of d'Alambertian produce Dirac delta-functions 
which will shrink one of the integration in the position space. To find the coefficients in front of the UD functions
we need to solve Bethe-Salpeter equation for this double-ghost correlator.  
Furthermore, in terms of this correlator all the correlators of dressed 
mean gluons can be expressed by using Slavnov-Taylor identity. Thus, we can conclude that the correlators of dressed mean 
fields in that theory which are off-shell correlators in the position space are   very complicated combinations of the 
three-point UD functions of space-time intervals.

\subsection*{Acknowledgments}

We are grateful to  Alvaro Vergara for his work with computer graphics for this paper.  
 The work of I.K. was supported by  Fondecyt (Chile) project \#1040368, 
and by Departamento de Ciencias B\'asicas de la Universidad  del B\'\i o-B\'\i o, Chill\'an (Chile). A.K. is supported by Fondecyt 
international cooperation project  \#7070064.

\end{document}